\documentclass{article}
\usepackage [pagewise, mathlines, displaymath]{lineno}
\usepackage[pdftex]{graphicx}
\usepackage{icrctc07}

\title{Radio detection of high-energy cosmic rays at the Pierre Auger Observatory}
\shorttitle{Radio detection at the Pierre Auger Observatory}
\authors{A.M. van den Berg$^{1}$, for the Pierre Auger Collaboration$^{2}$}
\shortauthors{Pierre Auger Collaborationl}
\afiliations{$^1$Kernfysisch Versneller Instituut, University of Groningen, NL-9747 AA, Groningen, The Netherlands \\
$^2$Av. San Mart\'{\i}n Norte 304, (5613) Malarg\"ue, Prov. de Mendoza, Argentina}
\email{berg@kvi.nl}

\abstract{
The southern Auger Observatory provides an excellent test bed to study the 
radio detection of extensive air showers as an alternative, cost-effective, and 
accurate tool for cosmic-ray physics. The data from the radio setup can be 
correlated with those from the well-calibrated baseline detectors of the Pierre 
Auger Observatory. Furthermore, human-induced radio noise levels at the 
southern Auger site are relatively low. We have started an R\&D program to 
test various radio-detection concepts. Our studies will reveal Radio Frequency 
Interferences (RFI) caused by natural effects such as day-night variations, 
thunderstorms, and by human-made disturbances. These RFI studies are 
conducted to optimise detection parameters such as antenna design, frequency 
interval, antenna spacing and signal processing. The data from our initial setups, 
which presently consist of typically 3 - 4 antennas, will be used to characterise 
the shower from radio signals and to optimise the initial concepts. Furthermore, 
the operation of a large detection array requires autonomous detector stations. 
The current design is aiming at stations with antennas for two polarisations, 
solar power, wireless communication, and local trigger logic. The results of 
this initial phase will provide an important stepping stone for the design of a few 
tens kilometers square engineering array.
}

\begin{document}
\maketitle

\section{Introduction}
The goal of the Pierre Auger Collaboration is to identify the origin of Ultra-High Energy (UHE) cosmic rays and to
unravel the mystery of the physics behind the cosmic acceleration mechanisms, leading to events in the Earths 
atmosphere with energies far beyond those obtained at any human-made accelerator. To achieve this goal, not only 
a precise energy measurement, but also a high angular resolution and the capability to determine the composition of 
rays ($\gamma$-rays, neutrinos or leptons, and hadrons) are of paramount importance. The baseline detector 
system of the Pierre Auger Observatory consists of an array of 1600 surface detectors (SDs)
complemented by 4
telescope buildings each housing 6 fluorescence detectors (FDs) \cite{daw2007}. Recently, the technique of detecting extensive air 
showers (EASs) with radio receivers has revived \cite{gre2003,fal2005,ard2005} after the initial measurements made in the 1960's  \cite{all1971}. 
Nowadays, there is more and more evidence that the underlying emission mechanism is geo-synchrotron radiation 
from electrons and positrons in the Earths atmosphere. Because close to the core, the thickness of an EAS induced by an UHE cosmic ray 
is less than about 5 m, this emission 
will be coherent if the detected frequency of the radio pulse is limited to about 100 MHz. For 
the detection of cosmic rays with an energy less than $10^{17}$ eV using radio techniques  
much progress has been made in recent years, especially because of the results obtained 
by the LOPES \cite{LOPESthisconference} and CODALEMA \cite{cha2007} collaborations. 

In principle, the advantage of radio compared to other systems used for the detection of UHE cosmic rays is large: 
radio signals are not absorbed nor deflected on their path, the amplitude of the signal is proportional to the primary energy of the 
incoming event, and one can study the shower front in detail. In addition, radio can have a 100\% duty cycle, which
is a factor of ten higher than fluorescence detection. Furthermore, the technique may provide additional information 
which is complementary to that from SD and FD, as it determines directly the evolution of the electromagnetic
properties of the shower in the atmosphere. 
This complementary information might open the possibility to study the composition of the primary event \cite{hugthisconference}. In addition, 
if one measures radio pulses with receivers distributed over a grid with a pitch size of many hundreds meters, a high 
angular resolution ($< 1^{\circ}$) for the arrival direction of the event can be obtained. 
The high duty cycle will provide us with 
many more events which are needed for a statistical analysis to determine possible anisotropies, to identify point 
sources, and to get a better insight into the composition of these UHE cosmic rays. 

Thus radio detection promises 
not only to be a bolometric measurement of the energy of the primary cosmic ray, but it can also give a precise 
determination of the arrival direction. Therefore, it might be a perfect additional tool for high-energy particle 
astronomy.

However, before radio detection of UHE cosmic rays can be regarded as a technique mature enough to deploy it over 
an area of the size of the Pierre Auger Observatories a substantial R\&D
program is required. This program will not only address the physics issues, but also the technological ones, and the 
investments and running costs. And it extends the continued efforts performed at the LOPES 
\cite{LOPESthisconference} and CODALEMA \cite{cha2007} sites in Europe. 

Initial investigations started in 2006 and we aim to merge our set ups with the foreseen infill-detector 
array at the southern Pierre Auger Observatory \cite{infill_heat,infill_amiga}. 
Starting in 2008, and for a period of 4 years we are planning to 
operate a 20 km$^{2}$ engineering array  
which will serve as a 
test bed to address these engineering and physics questions leading to the design of a many thousand kilometers 
square array.

\section{R\&D Program}

The aim of the present research and development program is to optimise the hardware and software required for an
area with a typical dimension of 20 km$^{2}$ which will be deployed at the southern site of the Pierre Auger Observatory. 
This array will provide its own trigger and read-out system and will thus operate independently from the baseline 
trigger and data-acquisition of the baseline detectors. The required correlation of events between the radio detection 
system and the baseline detectors will be performed using GPS time stamps. The optimisation of the whole read-out 
chain has various ingredients. It starts with the antenna design and operation, the preamplifier at the antenna station, 
the filters (analog and/or digital), the amplification, the receivers and digitisers, and the signal analysis.

\begin{figure}
\begin{center}
\includegraphics [width=0.48\textwidth]{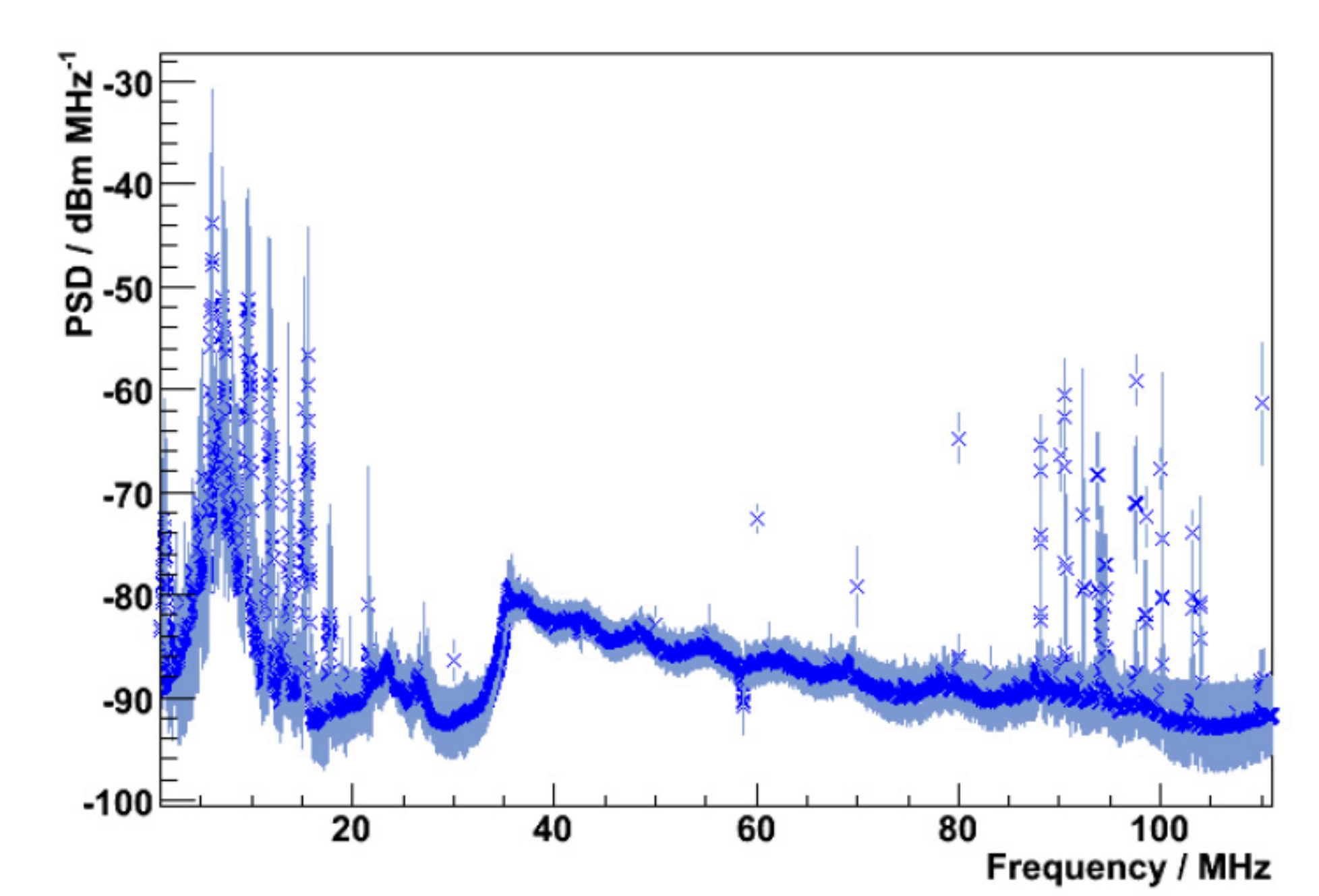}
\end{center}
\caption{
Spectral power density measured near the Balloon Launching Station using an LPDA and an
Anritsu MS2711D spectrum analyser.  
The antenna signal was amplified using a low-noise amplifier \cite{lopesstar} followed by
an RG213 cable with a length of 160~m.
}\label{fig1}
\end{figure}

The overall performance depends not only on a low system noise throughout the chain, but also on the capabilities 
for the suppression of Radio-Frequency Interferences (RFIs). These RFIs can be many fold: background from the 
Galactic sky, lightning from thunderstorms, carrier signals from radio and TV transmitters, and other human activities. 
It is known that strength of the man-made noise power can strongly depend on the site and that it rises 
steeply with decreasing frequency. Generally, the spectral power $S$ integrated over the solid angle measured by one 
 polarisation direction of a simple antenna with an effective aperture $A$ can be expressed as: 
\begin{eqnarray}\label{equ1} \nonumber
S_\nu &=& \frac{dP}{d\nu} = \frac{ \int I_{\nu} A d\Omega}{d\nu} \\ 
    &=&\frac{\int \left( I_{\nu,man} \!+\! I_{\nu,sky} \right) A d\Omega }{d\nu}   \nonumber
\end{eqnarray}
Here, the intensity $I_{\nu}$ of the induced power is written as the sum of the sky noise (mainly from Galactic and atmospheric origin) and 
noise induced by man-made sources. Both contributions are smoothly varying functions of the frequency $\nu$, though 
both intensities increase going to lower frequencies. In quiet rural areas $I_{\nu,sky}$ 
dominates $I_{\nu,man}$ for $\nu > $ 10 MHz.  
In addition to these smoothly varying contributions, we have to take into account the influence of strong transmitters 
used for radio and TV broadcasting. Finally, atmospheric 
noise can play an important influence on the ambient noise levels of the radio receivers. Atmospheric noise is 
caused by lightning, which at very low frequencies emits with a broad spectrum. A measurement of the ambient 
noise is shown in Figure \ref{fig1}. Transient noise, which can strongly fluctuate in amplitude and rate, can produce
signals which look very much like those of cosmic rays; therefore,  the reduction of this noise has to studied carefully.

The antennas which are being used in the present R\&D program are various types of dipole antennas. One system
uses a dual active fat dipole antenna (AFDA) which has a total length of 1.2 m mounted on a post of 1.0 m height, 
where two polarisations are measured. The AFDAs have a rather flat antenna gain over a large band 
\cite{cha2007}. 
Another design presently being tested, is a relatively large dual logarithmic periodic dipole antenna (LPDA). In this 
case, each antenna station has on one pole at a height of about 4 m two antennas, one for each polarisation. The 
antenna gain for these LPDAs drops relatively fast below 35 MHz, and is rather flat in the band between 35 and 90 
MHz \cite{lopesstar}. 
Finally, we use a dual inverted V-shaped dipole antenna (IVDA) \cite{fal2005}, also mounted on a post at a height of 5 m.
This antenna has a smoothly varying gain with a maximum around 60 MHz. 

\begin{table*}
\begin{center}
\begin{tabular}{|c|c|c|c|c|c|c|c|c|}
\hline
system	&antenna	&antenna	&wireless	&total	&trigger	&pass filter	&DAQ	&dynamic	\\
		&type	&band width	&	&gain	&		&			&sampling	&range	\\
		&		&		&		&		&		&			&rate	&		\\
		&		&(MHz)	&		&(dB)	&		&(MHz)		&(MS s$^{-1}$)	&(bits)\\
\hline
1		&AFDA	&$0.1-200$&yes	&35		&SoT	&$0.1-90$		&250		&8		\\
2		&LPDA	&$35-90$	&yes		&$<$75	&SoT/T3		&$13-85$		&3 * 40		&10		\\
3		&LPDA	&$35-90$	&no		&55		&SoT	&$41-79$		&80		&12		\\
4		&LPDA	&$35-90$	&no		&41		&SoT/P	&$25-75$		&400		&12		\\
5		&IVDA	&$35-80$	&no		&41		&SoT/P	&$25-75$		&400		&12		\\
\hline
\end{tabular}\label{tablebig}
\caption{Configurations used in the present R\&D phase (see text for details).}
\end{center}
\end{table*}
Also for the readout chain, different techniques are being used. In all cases, low noise amplifiers are mounted at the 
front end of the antenna. In the present R\&D stage, the signals are transported over relatively short ($<10$ m) or long 
(160 m) cables. In the case where the cables are short, digitisation takes place close to the antenna station and the
 data are transferred by a wireless link to a nearby central system. Here, the power for the antenna stations is being 
 generated by solar power units. 
 The digitisers in this case are either a system based on the simultaneous 
 digitisation of 3 slightly overlapping frequency bands each with a sampling rate of 40 MS s$^{-1}$ and a dynamic range of 10 
 bits or a system which has a rather broad band with a width of 250 MS s$^{-1}$ and a dynamic range of 8 bits. For the 
 antennas which are connected by the relatively long cables to the data-acquisition system, bias-tees connected to 
 the available infrastructure in the Balloon Launching Station of the Observatory are used to power the pre-amplifiers. 
 For these antennas, digitisation systems with a relatively large dynamic range (12 bits) are being used. One of these 
 12-bit systems has a rate of 80 MS s$^{-1}$, the other one has a rate of 400 MS s$^{-1}$. Triggers for these systems 
 are based on signal-over-threshold (SoT) or using an external trigger. This external trigger is generated by an 
 external pulser (P) or by a trigger of the SD array (T3).  When an external pulser is used the ambient background 
 can be measured with these data-acquisition systems. To avoid false triggers, when operated in the 
 signal-over-threshold mode, pass filters are used to suppress the received power for the regions below about 25 
 MHz and beyond 80 MHz. 
 The power suppression is typically 80 to 100 dB. 
 Table 1 summarises the various parts used in the present stage of the R\&D program. 
 
To increase the number of events which can be observed with the 
baseline detectors, an infill SD station has been deployed near both sites used for our radio R\&D program, one site 
near the Balloon Launching Station, the other near the Central Laser Facility.  In this way the SD-trigger rate is 
increased by a factor of 20, yielding about 2 events with an energy larger than 0.3 EeV per day within a radius of 500 
m from our test sites. 

To evaluate the different systems, we define a signal-to-noise ratio ($R$) based on the power 
measured by each of the different systems: 
\begin{equation}\label{equ2}\nonumber
R = \frac{P_{max}(signal)}{\int_{\nu_1}^{\nu_2} S_\nu(noise)  d\nu} \nonumber
\end{equation}
where $P_{max}(signal)$ is the interpolated maximum power of a radio signal induced by an 
EAS in the frequency band 
between $\nu_1$ and $\nu_{2}$ and $S(noise)$ is the 
power of the noise spectrum integrated over the same bandwidth. The value of $R$ 
will be a leading factor in the decision for the design and construction of an engineering array of about 100 - 150 
antennas, which we plan to deploy in the engineering array. 
Antenna stations of this array have to 
operate on solar power and data transmission will be based on wireless communications. 

\section{Conclusions}
We have started an R\&D program for the detection of UHE cosmic rays within the Pierre Auger Observatory. 
Presently, our main aim is to optimise our signal-to-noise ratio, testing several antenna concepts and read-out 
systems. The results of this initial phase will be used to design and construct a larger array with a dimension of 
about 20 km$^{2}$.



\end{document}